\documentclass[12pt,a4paper]{article}
\usepackage[latin1]{inputenc}
\usepackage{amsmath}
\usepackage{amsfonts}
\usepackage{amssymb}
\usepackage{graphicx}
\usepackage{color}
\usepackage{natbib}

\usepackage{subcaption}

\author{Robert Cowell\\
  Faculty of Actuarial Science and Insurance\\
  Cass Business School\\
  City University London\\
  106 Bunhill Row, London EC1Y 8TZ, UK\\
  email: r.g.cowell{@}city.ac.uk
}

\title{Combining allele frequency uncertainty and population substructure corrections in forensic DNA calculations }
\date{}

\begin{document}

\maketitle
\begin{abstract}
In  forensic DNA calculations of relatedness of individuals and  in
DNA mixture analyses,  two  sources of uncertainty   are present concerning  the 
allele frequencies used for evaluating  genotype probabilities when evaluating likelihoods. They are: (i)  imprecision in the  estimates of the allele frequencies in the 
population  by using an inevitably finite database of DNA profiles to estimate 
them; and (ii) the existence of population substructure. 
\cite{green/mortera:aoas2009} showed that these effects may be taken into account \textit{individually}   using  a common Dirichlet model within a Bayesian network formulation, 
but that  when taken  \textit{in combination} this is not the case; however they suggested an approximation that could be used.
Here we develop a slightly different approximation that is shown to be exact in the case of a single individual. We demonstrate  the closeness of the approximation numerically using a published database of allele counts, and illustrate the effect of  incorporating the approximation into  calculations of a recently published  statistical model of  DNA mixtures.
\end{abstract}

\subsection*{Keywords} Genotype probabilities; uncertain allele frequency; population substructure; DNA mixtures.

\section{Introduction}

In a recent publication, \cite{cowell:etal:2015} presented a statistical model   for the quantitative peak information
obtained from an electropherogram of one or more forensic DNA samples. The model incorporates  stutter and dropout artefacts, and allows for   the presence of multiple unknown individuals contributing to the samples. Using likelihood maximization, the model can be used to compare hypothetical assumptions about the contributors to the DNA samples, and for deconvolution of the mixture samples to generate a set of  joint genotypes of hypothesized untyped contributors that is ranked by likelihood.

The model of \cite{cowell:etal:2015}  assumes that the set of alleles in the population entering the likelihood calculations are in Hardy-Weinberg equilibrium, that is, there is no population substructure. It also assumes that the population allele frequencies are precisely  known. Neither assumption is valid for real casework. 
In the discussion section to \cite{cowell:etal:2015}, several contributors pointed to the need to accommodate these issues. 
Of particular interest for this paper is  the contribution from  Green  and Mortera, and the contribution  from 
Tvedebrink, Eriksen and Morling. 

The comments from the latter contributors  are  presented in more detail in \citep{tvedebrink:etal:2015}, and deal with incorporating population substructure into the mixture calculations using the Balding-Nichols correction \citep{djb/ran:fsi}. They show that a Dirichlet-multinomial distribution may be incorporated into an extension of the Markov model of allele probabilities of \citep{cowell:etal:2015} in order 
to take account of the Balding-Nichols $\theta$ correction. 
 Green discussed ongoing work with Mortera, extending earlier work in \citep{green/mortera:aoas2009}, for  modelling the uncertainty in allele frequencies arising from using observed frequency counts for alleles in a (finite) database.  Curiously, this also leads to a Dirichlet-multinomial distribution with parameters depending on the total database size and a Dirichlet prior parameter for allele frequencies.  In particular,  the same extension of the Markov model 
presented by \cite{tvedebrink:etal:2015} for population substructure may be used instead to model the uncertain allele frequency (UAF) by reinterpreting the  Balding-Nichols $\theta$ parameter as a function of the database size. The common occurrence  of 
the Dirichlet-multinomial distribution for \textit{separately} modelling either population substructure or uncertainty in allele frequency was shown by \cite{green/mortera:aoas2009} in terms of their  Bayesian network model. They show that
\textit{in combination}  they do not 
follow a Dirichlet-multinomial distribution 
when there are three or more founding alleles for a locus,  but suggest a first order   additive approximation that could be used for combining the two sources of uncertainty with the Dirichlet-multinomial framework.

This paper revisits the approximation suggested by  \cite{green/mortera:aoas2009} for combining corrections for population substructure and uncertain allele frequency. We develop a closed form formula slightly different to their additive approximation that is exact for a single person and which we propose may be used as an approximation for problems involving more than one person. 
We examine  the numerical accuracy of the approximation using a published population database, and how it affects likelihoods in the mixture example examined in \citep{cowell:etal:2015}. We begin by summarizing the Dirichlet models for each source of uncertainty taken separately, and then consider them in combination.

\section{Dirichlet modelling of population substructure correction}

A commonly applied probability model  to take account of shared ancestry in a population  is the $\theta$ correction of
 \cite{djb/ran:fsi}.  In this model, the 
 distribution of alleles in the population  is assumed to be known.
To account for the co-ancestry of individuals, a small parameter $\theta$ is introduced which perturbs the genotype probabilities away from those obtained under   Hardy-Weinberg equilibrium.  For example,  if an  allele of type $a$ occurs
in the population  with probability $p_a$, then  under Hardy-Weinberg equilibrium the probability for a randomly selected individual having the homozygotic genotype $(a,a)$ would be $p_a^2$, but with the $\theta$ adjustment it is instead 
$p_a(\theta + (1-\theta)p_a$. If $\theta=0$ we recover the Hardy-Weinberg values $p_a^2$. Values of $\theta$ used in forensic calculations  are typically in the range  0.01-0.03.

More generally, for a given locus 
denote the allele frequencies in the population by the vector $\mathbf{p} = (p_1, p_2, \ldots, p_K)$ for the $K$ alleles $(A_1, A_2, \ldots, A_K)$.
The probability of randomly selecting one allele of type $A_k$ is $p_k$.
The probability of randomly selecting a second allele of the same type, given we have seen already selected it once, is
$$ \frac{\theta + (1-\theta)p_k}{1}$$
The probability of randomly selecting a third allele of the same type, given we
have selected two copies, is
$$ \frac{2\theta + (1-\theta)p_k}{1 + \theta}$$
In general the probability of seeing an $a_k$-th allele of this type given we have seen
$a_{k}-1$ of that type previously, is
$$ \frac{(a_k-1)\theta + (1-\theta)p_k}{(a_k-1)\theta +1 - \theta}.$$

Hence the probability of seeing $a_k$ alleles of type $A_k$ will be
$$ \prod_{j_k=1}^{a_k} \frac{(j_k-1)\theta + (1-\theta) p_k}{ (j_k-1)\theta + 1-\theta }$$ 
Taking through the factor of $\theta$ top and bottom (assuming that $\theta>0$), and writing $\phi = (1-\theta)/\theta$, this may be rewritten as
$$\prod_{j_k=1}^{a_k} \frac{j_k-1 + \phi p_k}{j_k-1 + \phi}$$
Finally, taken over the set of  alleles  in a set of $I$ genotypes denoted by $g$ which have a total of $a_k$ alleles of type $A_k$ on the locus,  this result extends to
\begin{equation}
 P(g\vert \mathbf{p}) = 2^h \frac{\prod_k \prod_{j_k=1}^{a_k}(j_k-1 + \phi p_k)}{\prod_{j=1}^{2I} (j-1 + \phi)},
 \label{eq:coan1}
\end{equation}
where $h$ the number of heterozygous genotypes amongst the $I$ genotypes $g$.

\section{Dirichlet modelling of  uncertain allele frequency}
\label{sec:dir}

A  Bayesian approach to dealing with  uncertainty in the population allele frequencies  $ \pi(\textbf{p}) $ is to treat them as random variables with a Dirichlet prior distribution:
\begin{align*}
\textbf{p} & = (p_1,p_2,\ldots,p_k) \sim \mbox{Dir}(\alpha_1,\alpha_2,\ldots,\alpha_K),\\
\pi(\textbf{p})&= \Gamma(\alpha)\prod_{i=1}^K\frac{p_i^{\alpha_i-1}}{\Gamma(\alpha_i)},
\end{align*}
where each $p_i \ge0$, $\sum p_i=1$ and $\alpha= \sum \alpha_i$.

The $\alpha_i$ are commonly taken to be the observed allele counts in a database 
(so for a database of $M$ alleles this would give $\sum \alpha_i = M$).  In this case
 $\widehat{p_i} = \alpha_i/M$ would be the proportion of alleles of types $A_i$ of the locus in the database, 
and 
$$ \mbox{Dir}(\alpha_1,\alpha_2,\ldots,\alpha_K)\equiv  \mbox{Dir}(M\widehat{p_1}, M\widehat{p_2}, \ldots, M \widehat{p_K} ).$$

An alternative is to give the  $\alpha_i$ values  representing a prior belief about the occurrence of alleles in the population 
before the database is observed.
Two common suggestions for the values of  the prior parameters  $\alpha_i$ are to set $\alpha_i = 1/K$, or to  set $\alpha_i=1$.
\cite{curran:buckleton:2011} argue in favour of setting each  $\alpha_i = 1/K$. 
The  observed allele counts $\textbf{m} = (m_1,m_2,\ldots, m_K)$ of  alleles in the database for the locus are used to update this prior, on the assumption that the alleles in the database are independent and multinomially distributed given $\textbf{p}$. This leads to a posterior density also of Dirichlet type:
\begin{align*}
\pi(\textbf{p} &\vert \textbf{m})= \mbox{Dir}(\alpha_1+ m_1,\alpha_2+m_2,\ldots,\alpha_K+m_K)\\
 &\equiv \mbox{Dir}(M\widehat{p_1}, M\widehat{p_2}, \ldots, M \widehat{p_K} )\\
\end{align*}
where now $M = \sum_i {\alpha_i+m_i}$ and $\widehat{p_i} = E(p_i\vert \textbf{m})$.

Whichever approach is used, we have a  distribution of alleles that takes into account  the 
observed alleles in the database of the form:
$$\pi(\textbf{p} \vert \textbf{m}) = \mbox{Dir}(M\widehat{p_1}, M\widehat{p_2}, \ldots, M \widehat{p_K} ).$$

Genotype probabilities  are then obtained by averaging over this distribution:
\begin{align}
P(g)   &=  \int P(g\vert \mathbf{p}) d\pi(\mathbf{p}\vert m) \nonumber\\
&=    2^h \int \prod_k p_k^{a_k} d\pi(\mathbf{p}\vert m) \nonumber\\
&=    2^h \frac{\Gamma(M)}{\Gamma(M+2I)}
\prod_k\frac{\Gamma (M\widehat{p_k}+a_k) }{\Gamma(M\widehat{p_k})} \nonumber\\
&=    2^h  \frac{\prod_k \prod_{j_k=1}^{a_k}(M\widehat{p_k}+j_k-1)}{\prod_{j=1}^{2I} (M+j-1)}
 \label{eq:post1}
\end{align}

Note that  the right-hand-side of \eqref{eq:post1} is the same
 as on the right-hand-side of \eqref{eq:coan1} if we identify $M \equiv \phi = (1-\theta)/\theta$, (so that $\theta \equiv 1/(M+1)$), as was pointed out by 
 \cite{green/mortera:aoas2009}.  

In other words, the use of  Bayesian averaging with a Dirichlet prior to take account of uncertainty in allele frequencies in the population arising from using estimates from a finite database, is numerically equivalent to 
a Balding-Nichols $\theta$ correction for co-ancestry on setting $\theta = 1/(M+1)$ and using the $\widehat{p_i}$ as if they are the true  population allele frequencies. If in addition we take each of the  $\alpha_i=0$ (for the Dirichlet prior before the database allele counts are incorporated), then $M$ is the size of the database and the $\widehat{p_i}$ are the observed  proportions of the alleles in the database.

\section{A notational aside}

Before proceeding to the   consideration of allele frequency uncertainty combined with
substructure correction, we  define the following function:
\begin{equation}
f(x;n) = \prod_{j=1}^n (j-1+x) \label{eq:fdef}
\end{equation}
 
Expanding $f(x;n)$ as a power series in $x$, we denote the coefficients of the power of $x^j$ by $c(j,n)$ 
so that $f(x;n)= \sum_{j=0}^n c(j,n)x^j$.  We define  $c(0,0)=1$,  and  for every integer $j>0$ we have that  $c(0,j) = 0$.
It is not hard to show  the following low-order expansions of $f(x;n)$ for $n$ values up to $n=6$:
\begin{align*}
f(x;0) & = 1\\
f(x,1) &= x \\
f(x;2) &= x+x^2\\
f(x;3) &= 2x+3x^2+x^3\\
f(x;4) &= 6x+11x^2+6x^3+x^4\\
f(x;5) &= 24x+50x^2+35x^3+10x^4+x^5\\
f(x;6) &= 120x+274x^2+225x^3+85x^4+15x^5+x^6\\
\end{align*}

\section{Combining UAF and $\theta$ corrections}

Our proposed method for combining UAF with $\theta$ correction for evaluating a joint genotypes probability  is to calculate the
joint genotype  probability using  the 
$\theta$ correction given allele frequencies $\mathbf{p}$, and then to integrate the result with respect to  a Dirichlet
for
 the population allele frequencies $\mathbf{p}$ in order to take account of  uncertainty in their population values.

Now given the allele frequencies, the genotype probability for the (joint) genotype $g$  (of one or more individuals)  taking into account co-ancestry is given by (\ref{eq:coan1}):

\begin{equation*}
 P(g\vert \mathbf{p}) = 2^h \frac{\prod_k \prod_{j_k=1}^{a_k}(j_k-1 + \phi p_k)}{\prod_{j=1}^{2I} (j-1 + \phi)}.
\end{equation*}

\begin{align}
P(g) = 
E[P( g\vert  \mathbf{p})]
&=  2^h \int \frac{\prod_k \prod_{j_k=1}^{a_k}(j_k-1 + \phi p_k)}{\prod_{j=1}^{2I} (j-1 + \phi)} d\pi(\mathbf{p})
\label{eq:post2}
\end{align}

Thus we need to evaluate the multiple integral

$$ E\left[ \frac{\prod_k \prod_{j_k=1}^{a_k}(j_k-1 + \phi p_k)}{\prod_{j=1}^{2I} (j-1 + \phi)}\right]=
\frac{\Gamma(\phi)}{\Gamma(\phi+2I)}
\int \prod_k \prod_{j_k=1}^{a_k}(j_k-1 + \phi p_k) d\pi(\mathbf{p}).$$

We may rewrite the above expectation in terms of the $f$ function introduced earlier:

\begin{align*}
E\left[ \frac{\prod_k \prod_{j_k=1}^{a_k}(j_k-1 + \phi p_k)}{\prod_{j=1}^{2I} (j-1 + \phi)}\right] &=
\frac{\Gamma(\phi)}{\Gamma(\phi+2I)} \int \prod_k f(\phi p_k, a_k) d\pi(\mathbf{p})\\
&= \frac{\Gamma(\phi)}{\Gamma(\phi+2I)} 
\int \prod_k \left(\sum_{j_k=0}^{a_k} c(j_k,a_k)(\phi p_k)^{j_k}\right)d\pi(\mathbf{p})
\end{align*}

In the case where the set of genotypes $g$ is  that of  a single person, 
  the product in the integral has just one term and is readily evaluated.  We consider separately the  two cases
of a homozygous individual and a heterozygous individual.

\subsubsection*{Homozygous individual}
If the individual is homozygous $(A_k, A_k)$, then the expectation involves the
integral of $f(\phi p_k, 2) = \phi p_k + (\phi p_k)^2$.  If we denote the Dirichlet prior by
$\mathbf{p} \sim \mbox{Dir}(s_1,s_2, \ldots, s_K)$ where $s_j = M\widehat{p_j}$ and $s = \sum s_j = M$, then the expectation is  given by:

\begin{align*}
\frac{\Gamma(\phi)}{\Gamma(\phi+2I)} 
\int \left(\phi p_k + (\phi p_k)^2\right)d\pi(\mathbf{p})&=
\frac{\Gamma(\phi)}{\Gamma(\phi+2I)} 
\left( \phi \frac{s_k}{s} + \phi^2 \frac{s_k(s_k+1)}{s(s+1)}\right) \\
&= \frac{1}{\phi(\phi+1)}\phi\frac{s_k}{s}\left(1 + \phi\frac{s_k+1}{s+1}\right)\\
&= \frac{s_k}{s(\phi+1)}\left(1 + \phi\frac{s_k+1}{s+1}\right)\\
\end{align*}
We now using $\phi = (1-\theta)/\theta$, so that $1+\phi = 1/\theta$, we  rewrite the last line as
\begin{align*}
\frac{s_k}{s(\phi+1)}\left(1 + \phi\frac{s_k+1}{s+1}\right)&=
\frac{s_k}{s}\left(\theta + (1-\theta)\frac{s_k+1}{s+1}\right)\\
&= \frac{s_k}{s}\left(\theta +(1-\theta)\frac{s_k}{s}\frac{s}{s+1}+
(1-\theta)\frac{1}{s+1}\right)
\end{align*}
Now define
$$\tilde{\theta} = \theta + \frac{1-\theta}{s+1}.$$ Then we have that 
$$1-\tilde{\theta} = (1-\theta)\frac{s}{s+1}$$ from which we obtain
$$
\frac{s_k}{s}\left(\theta +(1-\theta)\frac{s_k}{s}\frac{s}{s+1}+
(1-\theta)\frac{1}{s+1}\right)=
\frac{s_k}{s}\left(\tilde{\theta} + (1-\tilde{\theta})\frac{s_k}{s}\right)$$

That is, given a Dirichlet distribution  for allele frequencies $\mathbf{p}\sim \mbox{Dir}(s_1,s_2,\ldots,s_K)$ to represent UAF, and the
Balding-Nichols correction parameter $\theta$ to represent population substructure, then for a homozygous individual the probability associated with the genotype is the same as if using
point estimates $\widehat{p_k} = s_k/s$ for the probabilities and using a substructure correction
with a  modified  correction parameter $\tilde{\theta} = \theta + (1-\theta)/(s+1)$.

We shall now show the same is true if the individual is heterozygous.

\subsubsection*{Heterozygous individual}

In the case of a heterozygous individual, with genotype $(A_j,A_k)$ and $j \ne k$,  the 
integral will involve the product $2f(\phi p_j, 1) f(\phi p_k, 1) = 2\phi^2 p_jp_k$, thus: 

\begin{align*}
2\frac{\Gamma(\phi)}{\Gamma(\phi+2I)} \int  \phi^2 p_j p_k d\pi(\mathbf{p})&=
	\frac{2}{\phi(\phi+1)}\phi^2 \frac{s_j}{s}\frac{s_k}{s}\\
	&= \frac{2}{\phi+1}\phi\frac{s_j}{s}\frac{s_k}{s+1}\\
	&= \frac{2}{\phi+1}\phi\frac{s_j}{s}\frac{s_k}{s}\frac{s}{s+1}\\
	&= 2 \frac{s_j}{s}\frac{s_k}{s} \frac{\phi}{\phi+1}\frac{s}{s+1}\\
\end{align*}

Now again using $\phi/(1+\phi) = 1-\theta$, the last line may be rewritten:
$$2 \frac{s_j}{s}\frac{s_k}{s} \frac{\phi}{\phi+1}\frac{s}{s+1}=2 \frac{s_j}{s}\frac{s_k}{s} (1-\theta)\frac{s}{s+1}$$
If we again define $\tilde{\theta} = \theta + (1-\theta)/(s+1)$, then
$1-\tilde{\theta} = (1-\theta)s/(s+1)$, 
hence the genotype probability can be written as
$$2 \frac{s_j}{s}\frac{s_k}{s}(1-\tilde{\theta} )$$

This is the what we would obtain by taking the $\widehat{p_j}={s_j}/{s}$ as the population allele frequencies and applied a population substructure correction using the
transformed parameter $\tilde{\theta} = \theta + (1-\theta)/(s+1)$:

$$ \frac{s_j}{s}\frac{s_k}{s}(1-\tilde{\theta} )=2\widehat{p_j}\widehat{p_k}(1-\tilde{\theta}).$$

Thus we have  shown that for the case of a single person, the  genotype probability may be found when  
combining  both UAF and population substructure corrections, by modifying the $\theta$ parameter value to take account of the size of the database by the transformation
\begin{equation}
\tilde{\theta} = \theta + \frac{1-\theta}{s+1} \label{eq:thetran}
\end{equation}
and using $\tilde{\theta}$ from \eqref{eq:thetran}  in the Balding-Nichols correction
 in which   the mean of the population allele frequency Dirichlet posterior  is treated  as being the population allele frequencies.

If we set $\theta=0$ in \eqref{eq:thetran}, then we obtain $\tilde{\theta} = 1/(s+1)$, thus recovering the equivalence in 
Section~\ref{sec:dir} found by \cite{green/mortera:aoas2009}.

The above result does not extend to the case for two or more  persons, although \eqref{eq:thetran} may be used as an approximation. 
\cite{green/mortera:aoas2009} suggested   an alternative approximation for  large $s$ ($M$ in their notation) and small $\theta$
 which is ``additive on the scale $M^{-1} = \theta/(1-\theta)$'', that is,
$$1/\phi_{GM} = 1/\phi + 1/s.$$
This is equivalent to the transformation of $\theta$ given by
\begin{equation}
\theta_{GM} = \frac{1 + \theta(s-1)}{s+1-\theta}  = \theta + \frac{(1-\theta)^2}{s+1-\theta}\label{eq:gm}
\end{equation}

\section{Numerical investigation of approximation}
\label{sec:numinv}
\subsection{Single person}

We now illustrate the accuracy of  the transformation \eqref{eq:thetran}, using population data for Caucasians on the marker vWA taken from \cite{butler:etal:03} shown in Table~\ref{tab:vwa}. (Similar results to those obtained below may be obtained for other markers.) 

\begin{table}[ht]
\centering
\caption{Allele counts for a sample of US Caucasians for the marker vWA. These counts have been obtained by rescaling the normalized frequencies given in \cite{butler:etal:03} by the database size of $s=604$, and rounding the results to the nearest integer.\label{tab:vwa} }
\begin{tabular}{l|ccccccccc}
  \hline
  \hline
Allele &13 & 14 & 15 & 16 & 17 & 18 & 19 & 20 & 21 \\ 
Count&  1 & 57 & 67 & 121 & 170 & 121 & 63 & 3 & 1 \\ 
   \hline\hline
\end{tabular}
\end{table}

\begin{figure}
\centering
\begin{subfigure}[t]{0.45\textwidth}
\centering
\includegraphics[width=\textwidth]{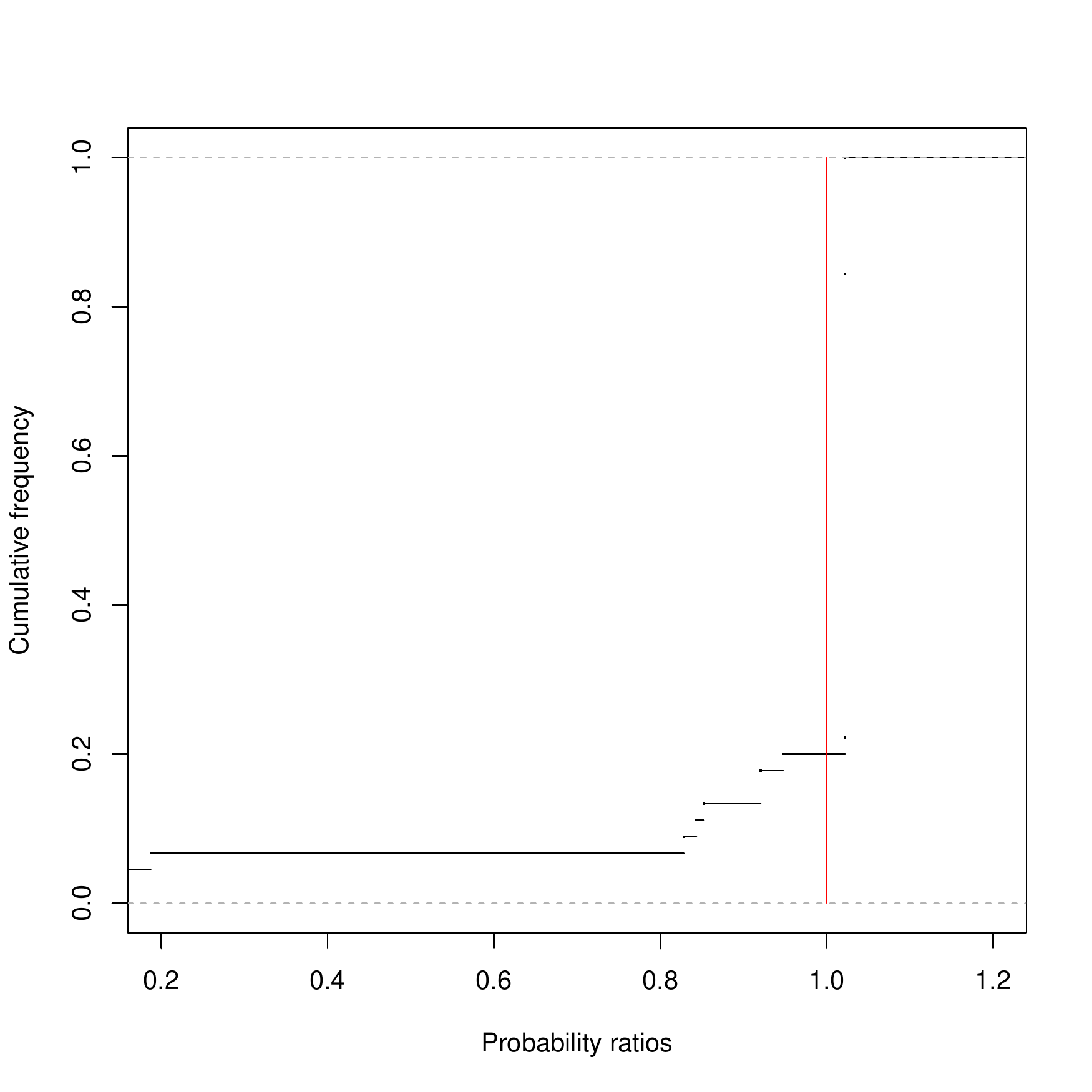}
\caption{Hardy-Weinberg}
\label{fig:hw1}
\end{subfigure}%
\hfill
\begin{subfigure}[t]{0.45\textwidth}
\centering
\includegraphics[width=\textwidth]{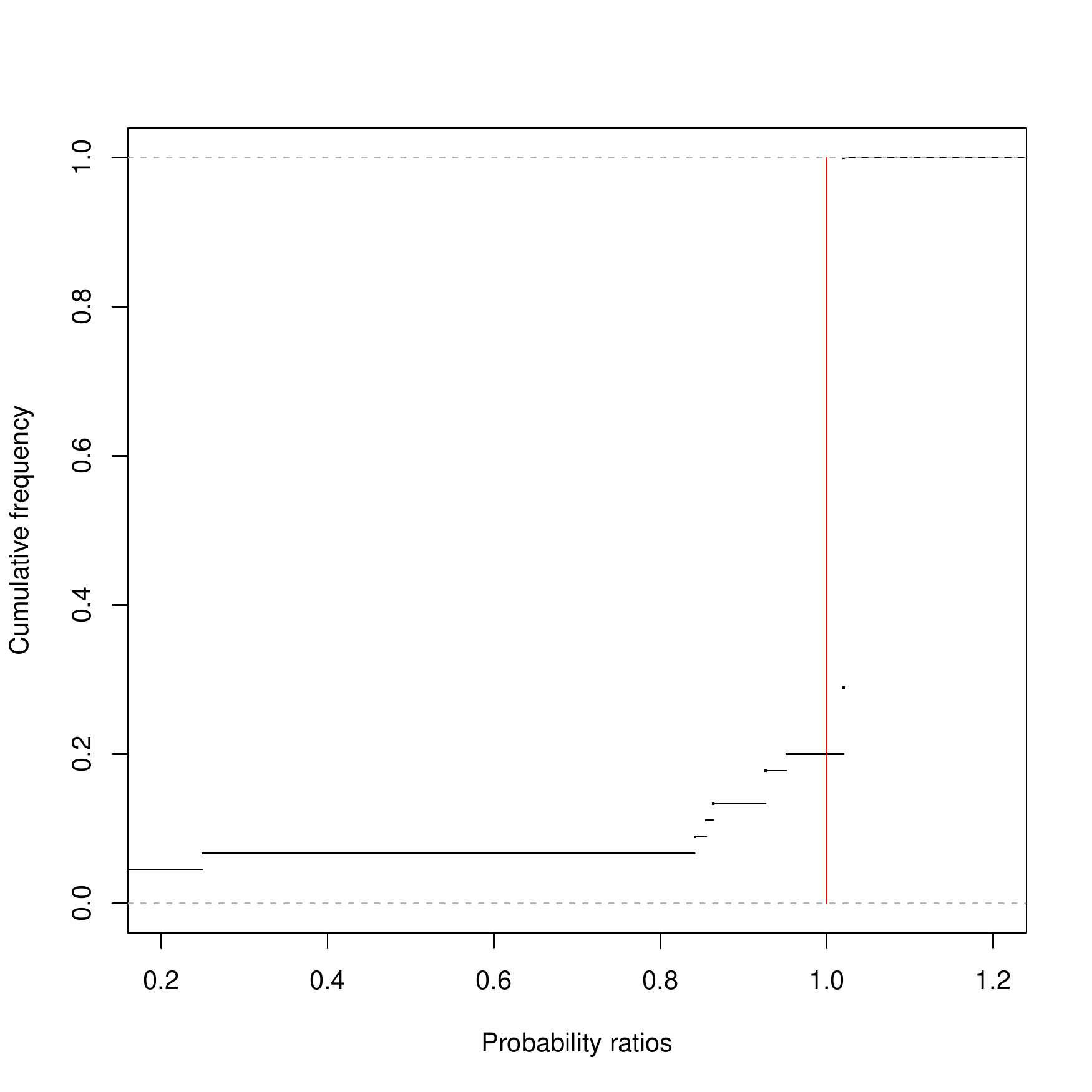}
\caption{UAF correction}
\label{fig:uaf1}
\end{subfigure}

\bigskip 

\begin{subfigure}[t]{0.45\textwidth}
\centering
\includegraphics[width=\textwidth]{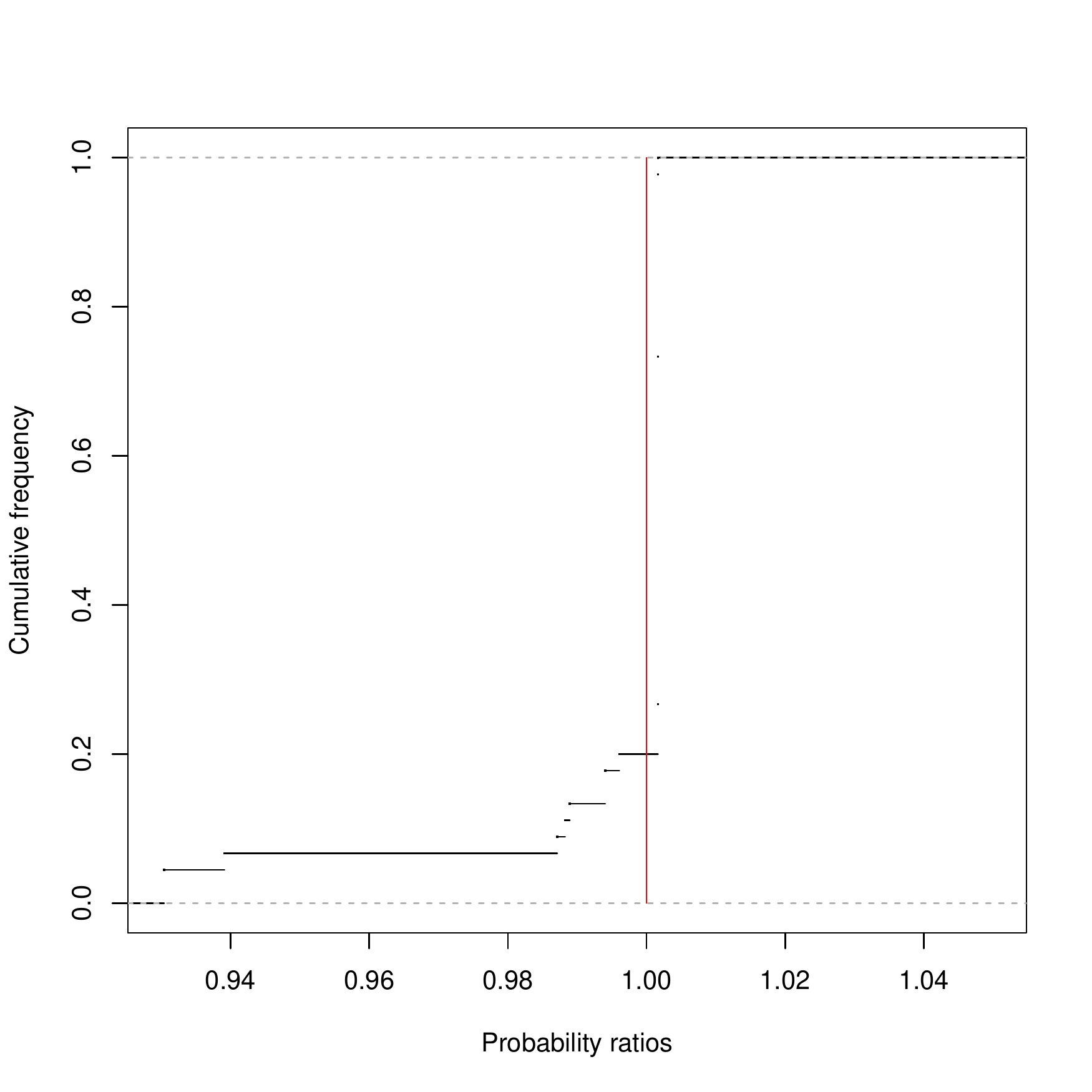}
\caption{Fst correction}
\label{fig:fst1}
\end{subfigure}%
\hfill
\begin{subfigure}[t]{0.45\textwidth}
\centering
\includegraphics[width=\textwidth]{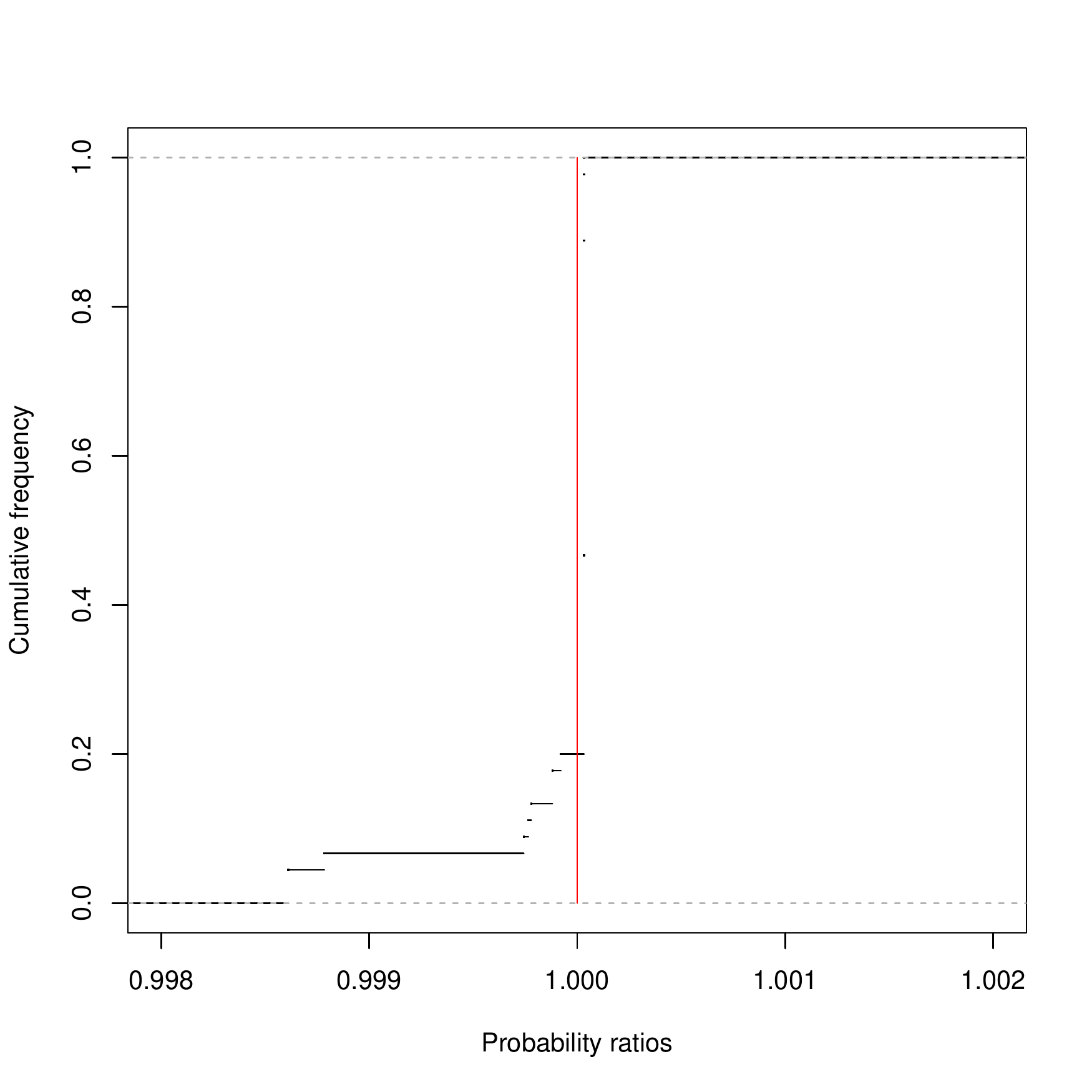}
\caption{Green-Mortera approximation}
\label{fig:gm1}
\end{subfigure}

\caption{Empirical cumulative distribution functions of the ratios to exact single person genotype probabilities  of corresponding genotype probabilities calculated 
under (a) Hardy-Weinberg equilibrium; (b) an uncertain allele frequency (finite database) correction without Fst correction; (c) Fst correction without finite database correction; (d) Green-Mortera approximation using finite database and Fst corrections. Note the much smaller range of ratios for plots (c) and (d). All plots are for the locus vWA. The database size is 604, and Fst correction where applied has $\theta = 0.02$.\label{fig:ecdf1}}
\end{figure}

We begin by looking at the distribution of the genotypes for a single individual. 
We do this by  comparing the distribution of genotype probabilities for a single person calculated under (a) Hardy-Weinberg equilibrium, (b) the correction for UAF only, (c) the $\theta$ substructure correction only, and (d) the Green-Mortera approximation
\eqref{eq:gm}. Each is compared against the exact correction for both  UAF and substructure in \eqref{eq:thetran} by calculating for each possible genotype the ratio of the probabilities under each of the approximations to the exact value.  Ideally the ratio should be 1. With the nine alleles of the vWA marker data in Table~\ref{tab:vwa} there are 45 possible genotypes. 
For the comparisons we take $\theta = 0.02$ and  $s=604$, the database size.  
In Figure~\ref{fig:ecdf1} are shown the
empirical cumulative distributions  of the ratio of the approximate to exact probabilities; the more the ratios are clustered around 1 the better the fit, as indicated by the vertical red lines. Note the smaller ranges for the subplots (c) and (especially) (d) compared to the subplots (a) and (b). In more detail,  the following ranges of ratios were found for the data of each plot: 
(a) $(0.0712, 1.0221)$;
(b) $(0.1422, 1.0204)$;
(c) $(0.9304, 1.0017)$;
(d) $(0.998611, 1.000033)$.

The plots indicate that the Green-Mortera approximation is excellent. This is  confirmed  by looking at the KL divergence between the approximate and exact genotype distributions; the values for the four approximations are respectively
(a) 0.001376; (b) 0.001127; (c) 6.411e-06; and (d) 2.537e-09.

\clearpage
\subsection{Two unrelated persons}

\begin{figure}
\centering
\begin{subfigure}[t]{0.45\textwidth}
\centering
\includegraphics[width=\textwidth]{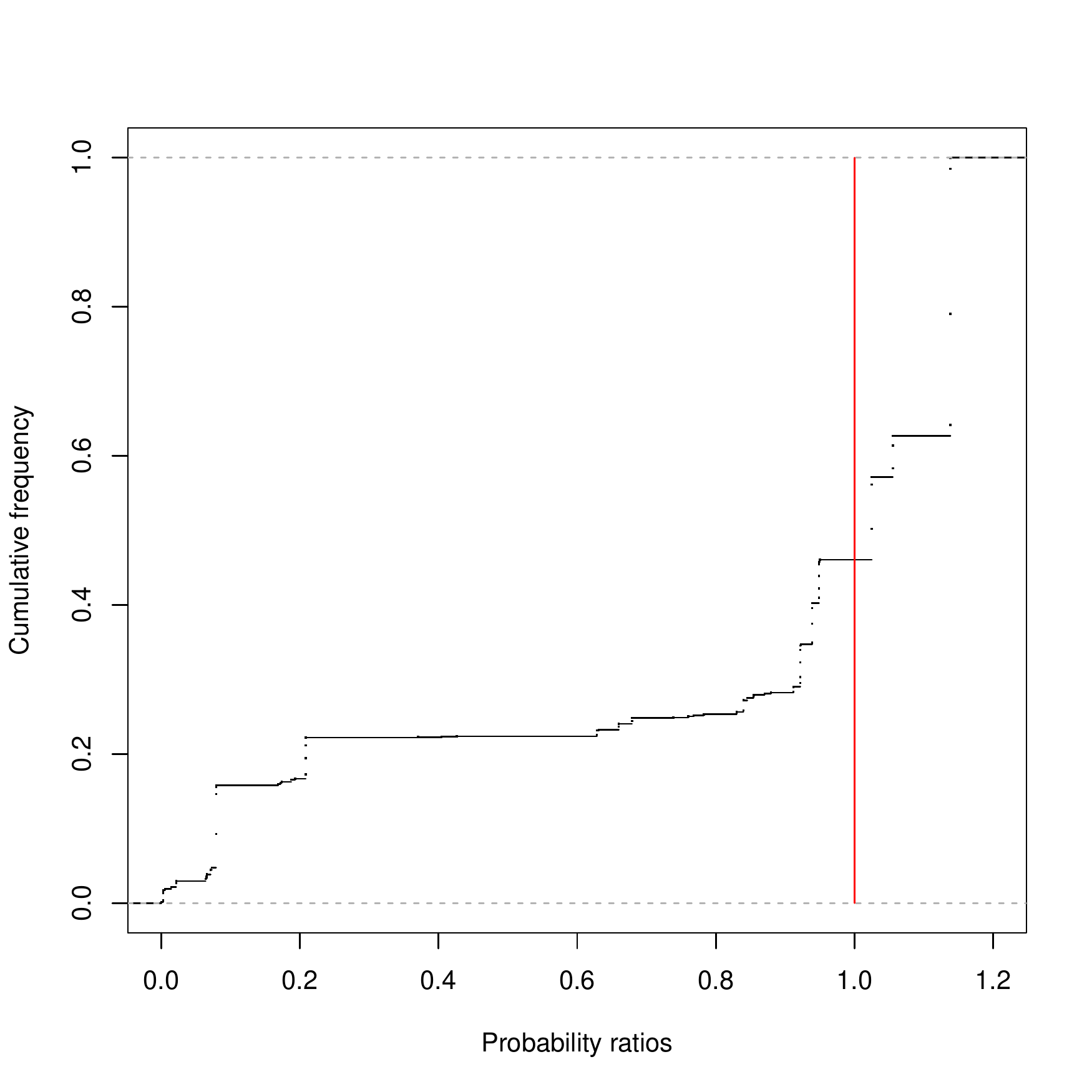}
\caption{Hardy-Weinberg}
\label{fig:hw2}
\end{subfigure}%
\hfill
\begin{subfigure}[t]{0.45\textwidth}
\centering
\includegraphics[width=\textwidth]{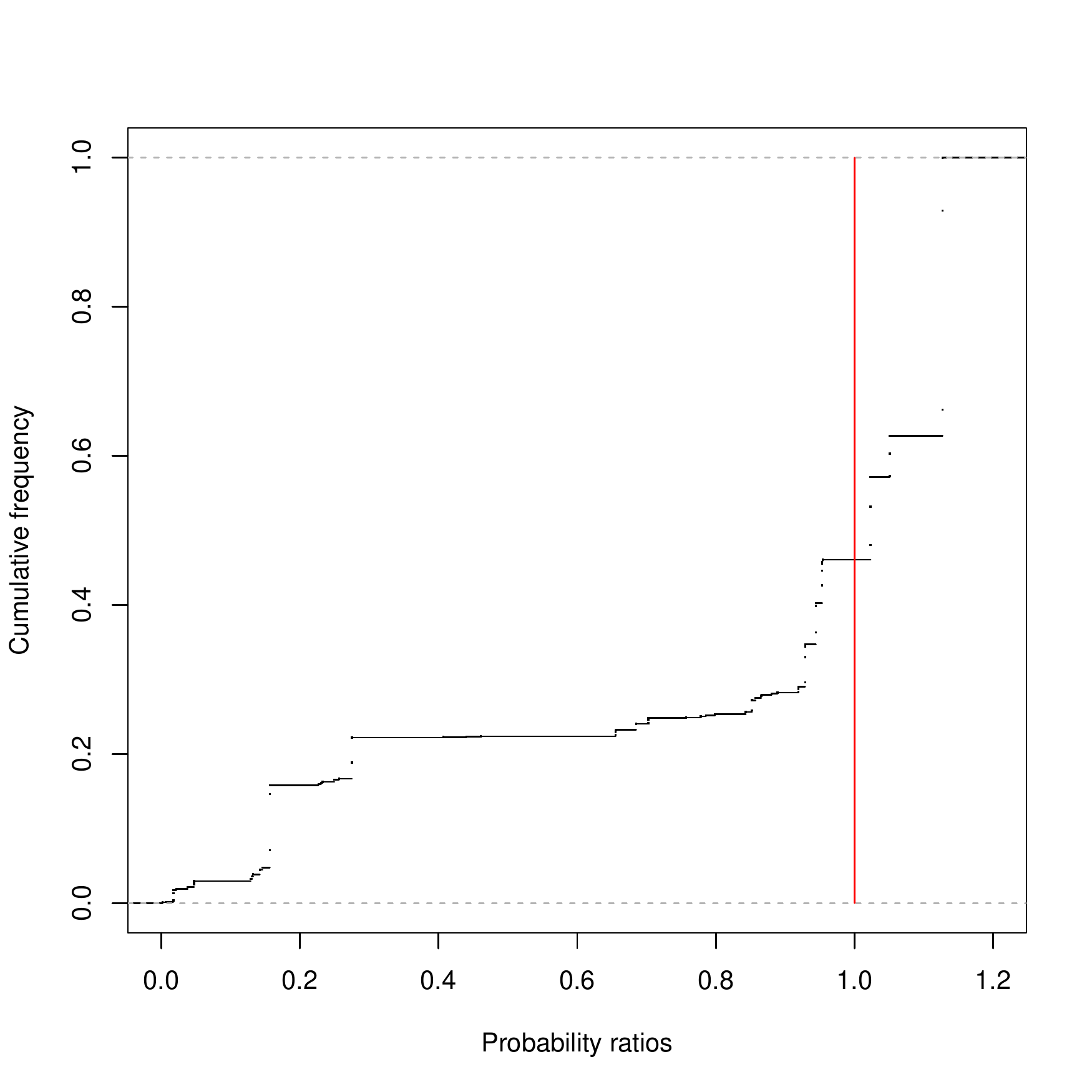}
\caption{UAF correction}
\label{fig:uaf2}
\end{subfigure}

\bigskip 

\begin{subfigure}[t]{0.45\textwidth}
\centering
\includegraphics[width=\textwidth]{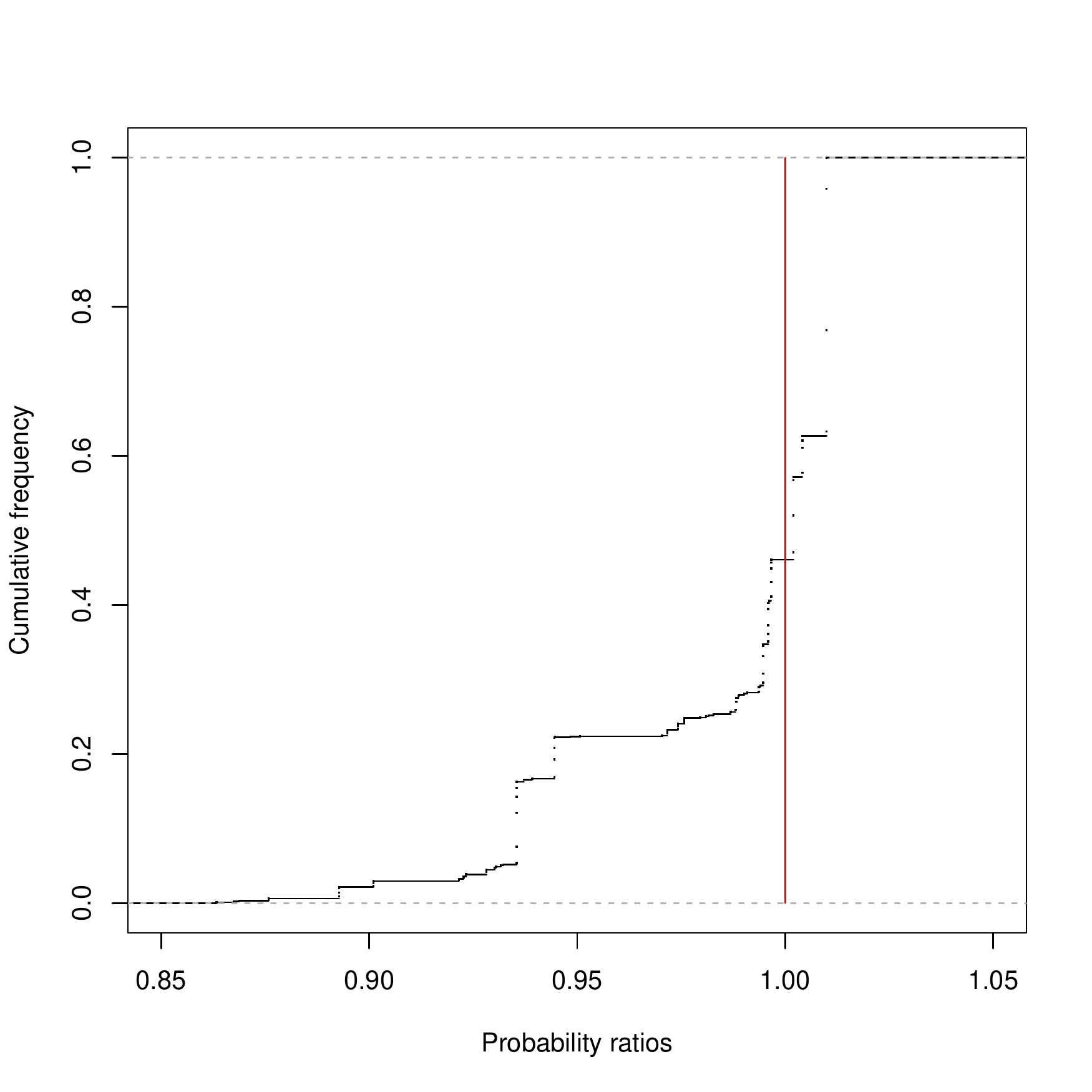}
\caption{Fst correction}
\label{fig:fst2}
\end{subfigure}%
\hfill
\begin{subfigure}[t]{0.45\textwidth}
\centering
\includegraphics[width=\textwidth]{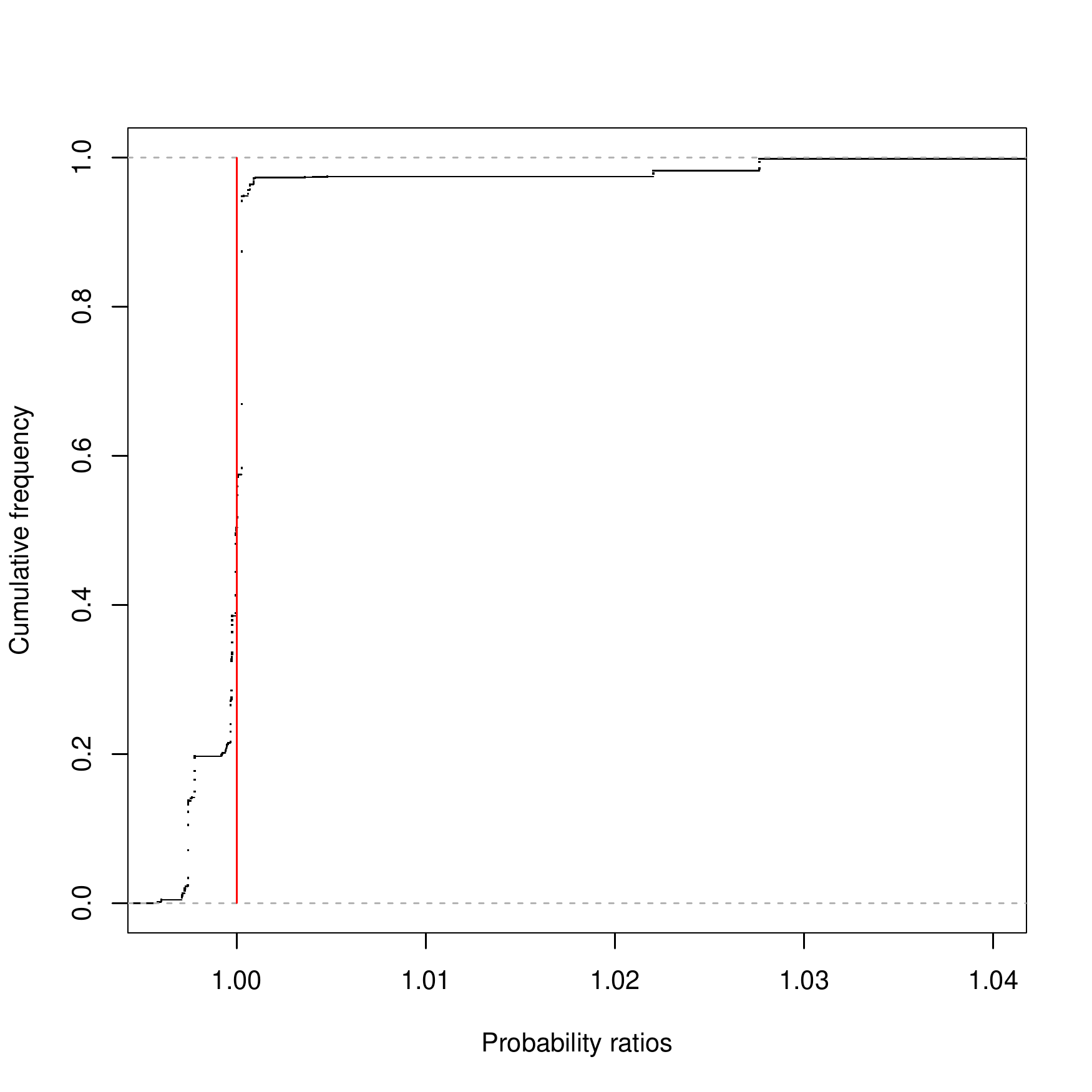}
\caption{Exact single-person approximation}
\label{fig:gm2}
\end{subfigure}

\caption{Empirical cumulative distribution functions of the ratios to the exact joint genotype probabilities of two unrelated 
individuals   of corresponding joint genotype probabilities calculated 
under (a) Hardy-Weinberg equilibrium; (b) an uncertain allele frequency (finite database) correction without Fst correction; (c) Fst correction without finite database correction; (d) approximation using exact adjusted theta value of
single person.  All plots are for the locus vWA. The database size is 604, and Fst correction where applied has $\theta = 0.02$.
\label{fig:ecdf2}}
\end{figure}

In Figure~\ref{fig:ecdf2} we compare the exact probabilities of the joint genotypes of two unrelated individuals (calculated with the aid of the $f(x;n)$ function in \eqref{eq:fdef}) to various alternatives. Specifically, the exact genotype distribution is compared to genotype frequencies assuming (a) Hardy-Weinberg equilibrium, (b) 
a correction for uncertain allele frequency alone, (c) a correction for population substructure alone, and (d) the approximation
using the exact single-person $\theta$ rescaling  \eqref{eq:thetran}. (A plot using the correction \eqref{eq:gm}  is very similar to (d) and is omitted.)
As in Figure~\ref{fig:ecdf1},  the plots in Figure~\ref{fig:ecdf2} show the
empirical cumulative distributions  of the ratio of the approximate to exact probabilities; the more the ratios are clustered around 1 the better the fit, as indicated by the vertical red lines. Again please note the smaller horizontal ranges for subplots (c) and (especially) (d) compared to plots (a) and (b). The following ranges of ratios of genotypes probabilities were found for the data of each plot: 
(a) $(0.0001, 1.1383)$;
(b) $(0.0018, 1.1271)$;
(c) $(0.8633, 1.01)$;
(d) $(0.995804, 1.07377)$.

The plots indicate that the rescaled $\theta$  approximation \eqref{eq:thetran}  is excellent. This can be confirmed numerically by looking at the KL divergence between the approximate and exact solutions; the values for the four approximations are:
(a) 0.008036; (b) 0.006552; (c) 3.6e-05; and (d) 4.9488e-08. (The KL divergence for the Green-Mortera approximation is 6.7575e-08, indicating the fit is not quite as good overall as that obtained using \eqref{eq:thetran}.)
Similar results may be found for the joint genotypes of three unrelated individuals, details are omitted here.

\clearpage

\begin{figure}[ht]
\begin{center}
\includegraphics[width=0.99\linewidth]{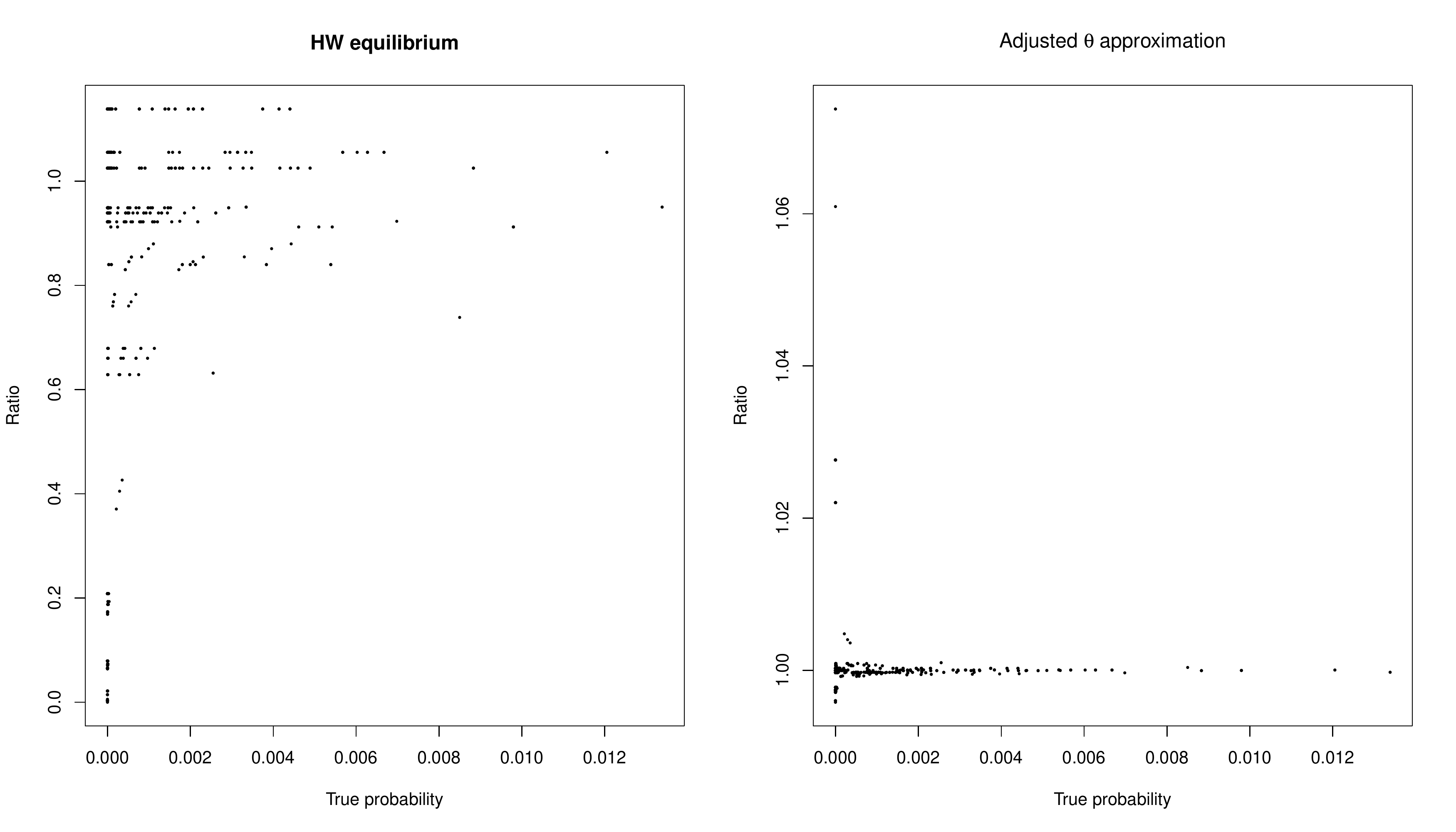}
\caption{Probability ratios  of each genotype possibility for two unrelated individuals calculated under HW equilibrium and 
adjusted $\theta$ correction, plotted against the true probability. Note the very different ranges of the vertical scales. \label{fig:scatter2} }
\end{center}
\end{figure}

Figure~\ref{fig:scatter2} shows scatterplots of the genotypes ratio probabilities against the exact probabilities for the Hardy-Weinberg and adjusted $\theta$ approximations for the case of two unrelated individuals - (note the very different vertical  scales). From this we see that the
genotype probability ratios that are furthest away from unity tend to be associated with those genotype combinations having lower probabilities.

\section{Application to mixtures}

We now examine the effect of applying the correction \eqref{eq:thetran} in  the analyses of DNA mixtures, specifically for the
example introduced by \cite{gill:etal:2008} and also analysed in \cite{cowell:etal:2015}. The reader is referred to these papers for full details concerning the example. Briefly, the example arose from casework in a murder investigation, in which two recovered bloodstain samples, labelled MC15 and MC18 were of importance. Analysis suggested that these DNA samples were each mixtures of DNA from at least three individuals. Three profiled individuals were of interest, the victim identified as $K_1$, an acquaintance of the victim identified as $K_2$, and the defendant, identified as $K_3$. 
A large part of the analysis carried out in \cite{cowell:etal:2015} assumed that the population allele frequencies were known, and that Hardy-Weinberg equilibrium was satisfied, although the latter assumption was relaxed for one specific scenario.  Here we revisit the evaluation of likelihoods using the model of \cite{cowell:etal:2015}, looking at the impact of taking into account UAF and population substructure corrections. Our analyses are based on the same US Caucasian data of \cite{butler:etal:03} used in Section~\ref{sec:numinv} and also used in \cite{cowell:etal:2015}.  As highlighted in the introduction,  \cite{tvedebrink:etal:2015} showed that the Balding-Nichols $\theta$  substructure correction  could be taken into account in the Markov network of \citep{cowell:etal:2015} used for computing  likelihoods, and as we have seen, setting $\theta =1/(s+1)$
means that the same computational framework can accommodate finite databases used for estimating allele frequencies in the population. To cake care of both UAF and $\theta$ correction, we use the approximation of \eqref{eq:thetran}.

We thus consider evaluating maximized log-likelihoods for the four sets of parameter values:
\begin{itemize}
\item HW: $\theta = 0$ and $s = \infty$, so that $\tilde{\theta} = 0$.
\item UAF: $\theta = 0$ and $s = 604$, so that $\tilde{\theta} = 0.0017$.
\item Fst: $\theta = 0.02$ and $s = \infty$, so that $\tilde{\theta} = 0.02$.
\item UAF+Fst: $\theta = 0.02$ and $s = 604$, so that $\tilde{\theta} = 0.0216$.
\end{itemize}

Table~\ref{tab:loglik} shows the maximized log-likelihoods for various
hypotheses regarding mixture contributors, obtained under the 
parameter settings above, for the peaks-height evidence given the
hypotheses. Note that the four values are the same on the first line
because we condition on the genotypes of the profiled hypothesized
contributors and there are no unknown 
profiles.  The possible prosecution hypotheses have the
defendant $K_3$ present as a contributor to the mixtures, defence
hypotheses do not.  However the profile of $K_3$ \textit{is} used in calculating the likelihoods of the
defence scenarios, as it (together with the profiles of $K_1$ and $K_2$)
will affect the genotype probabilities of the unknown contributors
\textit{unless} the alleles are in Hardy-Weinberg equilibrium and the
population allele frequencies are assumed known. This is a subtle but crucial
point that has implications for calculating likelihood ratios when
comparing prosecution and defence hypotheses, and is discussed further 
 in Section~\ref{sec:defcalc}.

\begin{table}[htbp]
\centering
\caption{Maximized log-likelihoods of three and four person scenarios obtained by analyzing the mixtures MC15 and MC18 both singly and together.\label{tab:loglik}}
\resizebox{\textwidth}{!}{%
\begin{tabular}{llcccc}
  \hline
Trace & Hypothesis & HW & UAF & Fst  & UAF+Fst \\ 
  \hline
  MC15 & $K_1+K_2+K_3$  & -118.087 & -118.087 & -118.087 & -118.087 \\ 
  MC15 & $K_1+K_2+U$ &	-130.201	&-130.117&	-129.492	&-129.452\\ 
  MC15 & $K_1+K_2+K_3+U$  & -118.043 & -118.044 & -118.051 & -118.051 \\ 
  MC15 & $K_1+K_2+U_1+U_2$ & -129.326&	-129.233	&-128.530	&-128.484 \\ 
  MC18 & $K_1+K_2+K_3$  & -130.148 & -130.148 & -130.148 & -130.148 \\ 
  MC18 & $K_1+K_2+U$ & -143.451	&-143.352&	-142.604	&-142.555\\ 
  MC18 & $K_1+K_2+K_3+U$  & -130.091 & -130.092 & -130.100 & -130.100 \\ 
  MC18 & $K_1+K_2+U_1+U_2$ & -143.361&	-143.262&	-142.521&	-142.473\\ 
  MC15 and MC18 & $K_1+K_2+K_3$  & -248.337 & -248.337 & -248.337 & -248.337 \\ 
  MC15 and MC18 & $K_1+K_2+U$ & -262.442	&-262.341	&-261.579&	-261.529 \\  
  MC15 and MC18 & $K_1+K_2+K_3+U$  & -248.214 & -248.215 & -248.232 & -248.233 \\ 
  MC15 and MC18 & $K_1+K_2+U_1+U_2$ & -262.296	&-262.198	&-261.457&	-261.409 \\ 
   \hline
\end{tabular}}
\end{table}

A prosecution and a defence hypothesis can be compared  by taking differences of their log-likelihoods to give a log-likelihood ratio in favour of the prosecution: six such combinations are shown in Table~\ref{tab:lograt}.  We see that in all six cases,  in taking into account  uncertainty in the allele frequencies and population substructure, the log-likelihood ratios  in  favour of the prosecution cases decrease as $\tilde{\theta}$ increases.  However this is but one casework mixture  example, and it should  not be inferred that this behaviour will always hold for other casework examples, although a heuristic argument is given in Section~\ref{sec:defcalc} that suggests this behaviour will be typical.

\begin{table}[htbp]
\centering
\caption{Log-likelihood ratios, in favour of the prosecution case of
  the presence of $K_3$ in the mixture(s), comparing several combinations
  of prosecution and defence hypotheses.\label{tab:lograt}}
\resizebox{\textwidth}{!}{%
\begin{tabular}{lccccc}
  \hline
Trace & Hypotheses & HW & UAF & Fst & UAF+Fst  \\
  \hline
MC15 & $K_1+K_2+K_3$  vs $K_1+K_2+U$ & 12.114	&12.030	&11.406	&11.365 \\ 
  MC15 & $K_1+K_2+K_3+U$  vs $K_1+K_2+U_1+U_2$ & 11.282&	11.189	&10.479&	10.433 \\ 
  MC18 & $K_1+K_2+K_3$  vs $K_1+K_2+U$ & 13.304&	13.205	&12.456	&12.408 \\ 
  MC18 & $K_1+K_2+K_3+U$  vs $K_1+K_2+U_1+U_2$ & 13.270	&13.170	&12.421&	12.372 \\ 
  MC15 and MC18 & $K_1+K_2+K_3$  vs $K_1+K_2+U$ & 14.104&	14.004	&13.241	&13.192 \\ 
  MC15 and MC18 & $K_1+K_2+K_3+U$  vs $K_1+K_2+U_1+U_2$ & 14.083	&13.983&	13.225&	13.175 \\ 
   \hline
\end{tabular}}
\end{table}

\section{Evaluating defence hypothesis likelihoods   for mixtures}
\label{sec:defcalc}
It is important to note that for the previous mixture examples, if we
did not include knowledge of the defendant's profile when calculating
likelihoods for the defence hypotheses, we would have found that the
likelihood ratios in favour of the prosecution cases would increase as
$\tilde{\theta}$ increases.  This is a subtle and crucial issue which
we expand upon here, using the two competing hypotheses of the first line of
Table~\ref{tab:lograt} for illustration.

If we denote the peak height evidence from the mixture by ${\cal E}$,
then the prosecution likelihood may be written as
$$L({\cal E}\vert  K_1,K_2, K_3 ,H_p) $$ 
where $H_p$ is the prosecution case that ``the three profiled individuals,   $K_1,K_2$ and $K_3$ all contributed to the mixture, and nobody else did".

The defence hypothesis amounts to replacing $K_3$ as a contributor to
the mixture with a random unrelated unknown person, denoted by $U$. The
defence likelihood may be written as
$$L({\cal E}\vert  K_1,K_2, K_3 ,H_d)  = \sum_{u}  L({\cal E}\vert  K_1,K_2, u, H_d)P(u\vert  K_1,K_2, K_3 ,H_d)$$
where the sum ranges over the possible genotypes of the unknown individual $U$, and  the defence hypothesis could be framed as ``$K_1, K_2$ and an unrelated individual $U$ contributed to the mixture \textit{and my client $K_3$ did not}".

Now we have assumed all the individuals are unrelated under both
hypotheses.  Under Hardy-Weinberg equilibrium and assuming the  allele frequencies
are known, ($\theta=0$), we have that $P(u\vert K_1,K_2, K_3, H_d ) =
P(u)$. However this equality does not hold if either of these
conditions is not valid, and one must retain all three profiled
individuals in the conditioning.  In particular, when evaluating the defence likelihood summation, 
it would be an error
to set $P(u\vert K_1,K_2, K_3 ,H_d) = P(u\vert K_1,K_2 , H_d)$ for each genotype $u$, even
though $K_3$ is not present in the mixture under the defence
hypothesis.  Although $K_3$ is assumed not to be in the mixture,
$K_3$'s profile is still required to evaluate the genotype probability
of the unknown individual for the likelihood evaluation. It would be
easy to overlook this point when the defence hypothesis is stated
solely in terms of assumed contributors to the mixture, for example in the form
``$K_1, K_2$ and an unrelated individual $U$ contributed to the mixture". This error
would lead to (incorrect) likelihood ratio values that could be 
detrimental to the defence hypothesis. For example, making this error
we find that the four values  in the first line of
Table~\ref{tab:lograt}  increase with increasing
$\tilde{\theta}$, taking the values $12.114, 12.143, 12.457$ and $
12.484 $ respectively.  This is an overstatement, in the value of
log-likelihood ratio in the final column, of $12.484 -11.365 =1.119$
in favour of the prosecution. (Similar results  occur for the other scenarios.)

A heuristic argument to explain the decreases observed in
Table~\ref{tab:lograt} is as follows.  The prosecution case is that $K_3$
contributed to the mixture, and the prosecution gives a large
likelihood ratio in favour of their case.  The defence case is that
someone else other that $K_3$ contributed to the mixture. However to
obtain the large likelihood ratio that the prosecution obtains, this
other unknown person may be expected to have a similar genetic profile
to the defendant $K_3$ in order to explain the features of the mixture that are not
explained by $K_1$ and $K_2$, and thus will have many alleles in
common with $K_3$.  For such possible genotypes $u$ of the unknown that are similar to
$K_3$, we would  expect that $P(u\vert K_1,K_2, K_3, H_d )
> P(u\vert K_1, K_2, H_d)$, if either there is population substructure
or the allele frequencies are not assumed known, since  if an
allele from $K_3$ is seen it increases the chance of seeing it again
in another individual randomly selected from the population. Thus, in
the weighted sum forming the defence hypothesis, greater  weight tends to be given
to the terms with  genotypes similar to $K_3$'s profile. Hence we would
expect the defence likelihood to increase over the $\tilde{\theta}=0$
value, and hence the likelihood ratio in favour of the prosecution to
decrease.

On the other hand, suppose we used $P(u\vert K_1, K_2, H_d)$
in evaluating the defence hypothesis. Then those genotypes similar to
$K_3$'s profile will tend to have a lower probability as
$\tilde{\theta}$ increases, because the frequencies of alleles in
$K_1$ and $K_2$'s profiles that are not in $K_3$'s profile will tend
to be greater, and because the sum of the allele probabilities is constrained to be 1,
this implies a reduction associated with the probabilities of the other alleles, in particular those that
$K_3$ has that are not shared with either
$K_1$ or $K_2$.  Hence the terms in the weighted sum for the defence
hypothesis likelihood will tend to receive less weight for those genotypes $u$
similar to $K_3$'s profile, so that the overall defence likelihood sum will decrease, leading
to an increase of the likelihood ratio in favour of the prosecution
case.

\section{Summary}

We propose \eqref{eq:thetran} as a simple way to combine uncertainties
both in allele frequencies arising from their estimation using a
finite database and the Balding-Nichols $\theta$ correction for
population substructure. The resulting genotype probabilities appear
to be very accurate, and are exact for the genotype of a single
individual. The effect on modifying genotype probabilities in
evaluating maximized likelihoods and likelihood ratios has been
demonstrated for scenarios in a complex mixture example. For computer
systems analyzing DNA mixtures that can currently take account of
Balding-Nichols $\theta$ correction, the computational overhead of
using the approximation to additionally include UAF is negligible.
This is also true for computer systems that evaluate likelihoods for
relationship problems, such as paternity testing, where such systems
already are able to take account of population substructure. We have
highlighted a subtle and important issue in evaluating defence hypothesis
likelihoods for mixtures in the presence of allele uncertainty or population
substructure that, if overlooked, could lead to errors  detrimental to a defence case.

\section*{Acknowledgements}
The author would like to thank Peter Green for his comments on an earlier version of this paper.


\begin{thebibliography}{7}
\providecommand{\natexlab}[1]{#1}
\providecommand{\url}[1]{\texttt{#1}}
\expandafter\ifx\csname urlstyle\endcsname\relax
  \providecommand{\doi}[1]{doi: #1}\else
  \providecommand{\doi}{doi: \begingroup \urlstyle{rm}\Url}\fi

\bibitem[Balding and Nichols(1994)]{djb/ran:fsi}
David~J. Balding and Richard~A. Nichols.
\newblock {DNA} profile match probability calculation: How to allow for
  population stratification, relatedness, database selection and single bands.
\newblock \emph{Forensic Science International}, 64:\penalty0 125--140, 1994.

\bibitem[Butler et~al.(2003)Butler, Schoske, Vallone, Redman, and
  Kline]{butler:etal:03}
John~M. Butler, Richard Schoske, Peter~M. Vallone, Janette~W. Redman, and
  Margaret~C. Kline.
\newblock Allele frequencies for 15 autosomal {STR} loci on {U.S.}\
  {C}aucasian, {A}frican {A}merican and {H}ispanic populations.
\newblock \emph{Journal of Forensic Sciences}, 48\penalty0 (4), 2003.
\newblock Available online at www.astm.org.

\bibitem[Cowell et~al.(2015)Cowell, Graversen, Lauritzen, and
  Mortera]{cowell:etal:2015}
R.~G. Cowell, T.~Graversen, S.~L. Lauritzen, and J.~Mortera.
\newblock Analysis of forensic {DNA} mixtures with artefacts (with discussion).
\newblock \emph{Journal of the Royal Statistical Society: Series C (Applied
  Statistics)}, 64\penalty0 (1):\penalty0 1--48, 2015.

\bibitem[Curran and Buckleton(2011)]{curran:buckleton:2011}
James~M Curran and John~S Buckleton.
\newblock An investigation into the performance of methods for adjusting for
  sampling uncertainty in {DNA} likelihood ratio calculations.
\newblock \emph{Forensic Science International: Genetics}, 5\penalty0
  (5):\penalty0 512--516, 2011.

\bibitem[Gill et~al.(2008)Gill, Curran, Neumann, Kirkham, Clayton, Whitaker,
  and Lambert]{gill:etal:2008}
Peter Gill, James Curran, Cedric Neumann, Amanda Kirkham, Tim Clayton, Jonathan
  Whitaker, and Jim Lambert.
\newblock Interpretation of complex {DNA} profiles using empirical models and a
  method to measure their robustness.
\newblock \emph{Forensic Science International:Genetics}, 2:\penalty0 91--103,
  2008.

\bibitem[Green and Mortera(2009)]{green/mortera:aoas2009}
Peter~J. Green and Julia Mortera.
\newblock Sensitivity of inferences in forensic genetics to assumptions about
  founding genes.
\newblock \emph{Annals of Applied Statistics}, 3\penalty0 (2):\penalty0
  731--763, 2009.

\bibitem[Tvedebrink et~al.(2015)Tvedebrink, Eriksen, and
  Morling]{tvedebrink:etal:2015}
Torben Tvedebrink, Poul~Svante Eriksen, and Niels Morling.
\newblock The multivariate {D}irichlet-multinomial distribution and its
  application in forensic genetics to adjust for subpopulation effects using
  the $\theta$-correction.
\newblock \emph{Theoretical Population Biology}, 2015.

\end{thebibliography}
\end{document}